\documentclass[epsf,12pt]{article}  

\begin{document}


\title{Heavy Quarks:\\ Effective Theories, Lattice and Models \footnote{Summary Talk of the Heavy Quark Session at the 5th International Conference on Quark Confinement and the Hadron Spectrum, Gargnano, Brescia, Italy, 10-14 Sep 2002.}}

\author{J. Soto \\ \\ \small{\it Universitat de Barcelona} \\ \small{\it Departament d'Estructura i Constituents de la Mat\`eria}\\
\small{\it Diagonal 647 }, \small{\it   
08028 Barcelona}, \small{\it Catalonia, Spain} \\ 
\small{\it E-mail: soto@ecm.ub.es }}

\maketitle
\begin{abstract}

Heavy-light mesons, heavy quarkonium and doubly heavy baryons are briefly discussed. Effective field theories (EFTs) of QCD based on 
the heavy quark mass expansion $1/m_Q$  provide a unified framework to describe all three systems. They produce a number of model-independent results and 
simplify non-perturbative (lattice) calculations. Nevertheless, models are still very useful to obtain inexpensive estimates of many observables.
 It is emphasized that certain models can be accommodated in the EFT framework,
like
 non-relativistic potential models for heavy quarkonium. It is also outlined how to build a suitable EFT for doubly heavy baryons which might also accommodate 
existing models for these systems. Finally, a few lattice calculations which are badly needed as inputs for EFTs are pointed out. 

\end{abstract}

\section{Introduction}

Heavy quarks means charm, bottom and top in the Standard Model (SM). Their masses $m_c\sim 1.5$ GeV , $m_b\sim 5$ GeV and 
$m_t\sim 190$ GeV are considerably larger than  those of the so called light quarks, namely up, down and strange, which lie  
in the few MeV range. Furthermore, the masses of the heavy quarks turn out to be larger 
than the masses of the first gapped excitations 
made out of light quarks only ($\rho$, $N$, ..), which set the scale $\Lambda_{QCD}$. Most of the theoretical progress on the heavy quark physics
of QCD in the last 16 years emanates from exploiting the inequality $m_Q >> \Lambda_{QCD}$, a job which 
EFT techniques have proven to be extremely useful at. Indeed, suitable EFTs have been put forward for systems composed of a
single heavy quark ( Heavy Quark Effective Theory (HQET) \cite{hqet,isgurwise}) and for systems composed of a heavy quark and a heavy antiquark 
(Non-relativistic QCD (NRQCD)\cite{nrqcd}), which nowadays are well established. EFTs however are limited by the fact that
they do not address the dynamical properties related to the scale $\Lambda_{QCD}$, which require non-perturbative techniques,
but only parameterize them in a way consistent with QCD and the relevant kinematical situation. Hence they have to be supplemented
with additional non-perturbative QCD calculations, most of which are carried out by means of lattice simulations. 
Unfortunately, many times lattice results for real QCD (with three light fermions) are either not available or not accurate enough. Then detailed information 
about these systems can only be obtained through models.

Except for the contribution \cite{weber}, where general properties of relativistic bound state equations are discussed, the remaining contributions in the Heavy Quarks session were related to the following types of hadrons: (i) mesons composed of a single heavy quark (heavy-light mesons
$Q\bar q$ )
\cite{ebert,green,pham}, (ii)
mesons composed of a heavy quark and a heavy anti-quark (heavy quarkonium $Q\bar Q$) \cite{tseng,eidemueller,jain,zhu,graham,kinoshita,godfrey} and (iii) baryons composed of two heavy quarks
(doubly heavy baryons $QQq$) \cite{ebert2}.

\section{$Q\bar q$ systems }

HQET has become the standard tool to analyze these systems 
\cite{uraltsev}. HQET renders the dependence on the 
heavy quark mass explicit, but it does not address the dynamics of the light quark and accompanying gluons. Hence  
statements beyond those which follow from the spin and flavor symmetries~\cite{isgurwise} that HQET enjoys usually 
require external information. This information is mainly extracted from two sources.

The first source is modeling. Models are usually able to produce definite predictions on many observables and hence have the
 advantage that they can be tested against experimental results or serve experimentalists as a guidance on where to look at.
 We had a nice example of a relativistic quark model in the session \cite{ebert}. Unfortunately,  since the relation of models for heavy-light systems to
 fundamental QCD is at the present unclear, their failures do not
imply a failure of QCD and conversely  their successes do not imply that QCD works either.

The second source is lattice QCD. This is a fundamental approach, which in the past had difficulties to properly account for
chiral symmetry and to include
 light dynamical quarks. The first problem has been overcome in recent years  
(see \cite{latticechiral} and references therein) and progress is steadily been made
 in the latter. We had an interesting contribution \cite{green} where not only the spectrum of heavy-light mesons was studied in
the static limit with light dynamical quarks but also the so called radial distributions, which might be used as QCD inputs for suitable kernels of relativistic
bound state equation in certain models.

Weak decays of heavy-light mesons, in particular of $B$-mesons, are receiving a lot of both theoretical and experimental attention.
They are crucial to fix some of the $CKM$ matrix elements and hence to give a definite answer to the  long standing question whether the  
SM accounts for all $CP$ violation observed in nature. We had an interesting contribution~\cite{pham} where $CP$ violation was studied through 
asymmetries in non-leptonic three body $B$ decays. Non-leptonic $B$ decays usually require factorization hypothesis for weak matrix elements between 
hadronic states which sometimes are not in a solid footing. Progress in this direction has also been made in the last years 
(see for instance \cite{nb}) and the introduction of EFTs techniques to address this issue \cite{scet} appears very promising (see \cite{scetnew} for recent work and references).

\section{  $Q\bar Q$ systems}   

In the same way that HQET has become the standard theoretical tool to study heavy-light systems, NRQCD has done so for heavy quarkonium \cite{bodwin}.   
Although the lagrangian of NRQCD is nothing but the addition of the HQET lagrangians for a heavy quark and a heavy antiquark
plus local four fermion quark-antiquark interaction terms, the heavy quark mass dependence of the spectrum cannot be easily 
extracted from it, in contrast to heavy-light systems. This is due to the fact that dynamically generated scales like the 
typical momentum transfer $k$ (inverse size of the system) or the binding energy $E$ depend now on the heavy quark mass.

Before the introduction of NRQCD much work had been done, and is still done, using potential models. Writing down a
Schr\"odinger equation with a suitable potential appears to be a reasonable thing to do for two heavy quarks, and a considerable effort had been done in the past to obtain suitable potentials from QCD, both within \cite{potpert}  and 
beyond \cite{potential} perturbation theory. Nevertheless, it has not become clear until recently what is the exact relation of such models with NRQCD, and hence with QCD.

A key observation to establish this link was that NRQCD contains degrees of freedom 
 which are irrelevant for the kinematical situation of most of the heavy quarkonium states \cite{mont}.
It is then convenient to introduce a new EFT, which was called potential NRQCD (pNRQCD) \cite{mont,long}, where the irrelevant degrees of freedom are integrated out \cite{vairo}. 
At leading order pNRQCD reduces to a Sch\"odinger equation, although the aspect (explicit degrees of freedom) of the pNRQCD lagrangian beyond leading order
depends on the interplay between $k$ , $E$ and $\Lambda_{QCD}$. We will use this interplay
to classify heavy quarkonium states in nature and the various contributions to 
this session.

The situation  $k >> \Lambda_{QCD}$, $E$ corresponds to the weak coupling regime. In nature the $\Upsilon (1s)$, $\eta_b (1s)$
and would-be toponium resonances are in this situation. These states are in a very good theoretical control since NNLL \cite{pineda}
(see also \cite{iainandre} and references therein) and NNNLO \cite{hamburg} calculations have recently become available.
It is important to keep in mind that the non-perturbative effects in this 
situation are conveniently parameterized by condensates (local \cite{voloshinleutwyler} or non-local \cite{dosch,long} ) and not 
by introducing non-perturbative contributions to the potential (i.e. the typical linear term). In fact, it would be very important in order to 
reduce the systematic error of the above calculations to have up dated lattice estimates of these condensates. In any case, 
 precision experimental data on $\eta_b (1s)$ would be very welcome to check these calculations against nature. It is
then a good new the $\eta_b (1s)$ event found at CDF  which was reported in this section \cite{tseng}. 
The sum rule techniques reported in \cite{eidemueller}  are also related to the weak coupling regime. They allow the extraction of 
$\overline{\rm MS}$ charm and bottom masses from charmonium and bottomonium systems. We also had a report on a search
for top-antitop resonances at D0 \cite{jain}. This search is, unfortunately, not sensible to the  would-be toponium
resonances \cite{top} which will have to wait for the Next Linear Collider being built \cite{supriya}.   

The situation $k \sim \Lambda_{QCD} >> E$ corresponds to the strong coupling regime. In nature this situation appears to be reasonable
for all states below threshold which are not too close to it (except for the ones in the weak coupling regime mentioned above). 
Hence most of the experimental results are related to them, in particular three of
the experimental contributions to this session: the BES results reported in \cite{zhu}, which include a partial wave analysis of
radiative two pion and two kaon decays relevant to light meson spectroscopy, the Fermilab E835 measurement of the two photon 
width of ${\chi_c}_0$ \cite{graham}, and the results from BELLE \cite{kinoshita} which include a puzzling enhancement of double 
charmonium production (See \cite{blb} for a possible explanation). We also had an important theoretical contribution \cite{godfrey} with estimates on branching ratios of 
various production and decay channels for the missing quarkonium states below threshold ($\eta_b$, $h_b$ 
, $h_c$ and $D$-wave states). Let me make a few remarks concerning this contribution, which also apply to any potential model 
calculation. Potential models are indeed EFTs of QCD in this kinematical situation \cite{long} provided that: (i) the interaction with pseudo-Goldstone bosons, which is small, is neglected, (ii) the potentials 
in the model are the ones obtained from QCD \cite{pinedavairo}, or a good approximation to them (see for instance \cite{gunar}),
and  (iii) the usual formulas for the decay widths are corrected  in order to properly take into account the contributions of the 
NRQCD color octet operators \cite{pwave}. Similar modifications are expected in the production formulas but they have not 
been worked out yet. It would be very useful for eventual tests of QCD in heavy quarkonium physics that potential model practitioners 
take into account these facts in their calculations.   

\section{\bf $QQq$ systems}

In contrast to $Q\bar q$ and  $Q\bar Q$ systems,  no specific EFT has been built for doubly heavy baryons (yet). 
Hence, almost all theoretical information available on them is based on models, from which we had a nice example in this session 
\cite{ebert2}. It is interesting to elucidate how the missing EFT for $QQq$ systems  would look like \cite{kiselev}. Following 
the ideas of  \cite{mont} we should first integrate out degrees of freedom with energies at the scale $m_Q$ or higher. We would get
 the HQET lagrangian plus four fermion quark-quark local interaction terms, which would be nothing but NRQCD in the heavy quark-quark
 sector.
This  lagrangian can either 
 be used to produce lattice results, analogously to $Q\bar Q$ systems (see for instance \cite{latticeNRQCD}), or as a starting point of further EFTs
which, hopefully, would be closer to available models, like pNRQCD for $Q\bar Q$ systems is to potential models. In the weak 
coupling regime the $Q$-$Q$ 
interaction has an attractive channel and a repulsive one (like the $Q$-$\bar Q$ interaction) . Hence for very large $m_Q$ the
attractive channel would give rise to Coulomb-type bound states between the two quarks of energy $E\sim m_Q \alpha_s^2$ . If
  $m_Q \alpha_s^2 >> \Lambda_{QCD}$,  one could built an EFT for the $Q$-$Q$ pair analogous to pNRQCD in the weak coupling
 regime \cite{long}. 
The interaction of each Coulomb-type bound state with the non-perturbative gluons would produce
a HQET-like spectrum for the remaining interaction with the light quark \cite{sw}. In the strong coupling regime, namely when the typical 
momentum transfer between the heavy quarks $\sim  \Lambda_{QCD}$, the light quark is expected to affect the leading interaction 
between the heavy quarks in a similar way as the so called hybrid potentials for $Q\bar Q$ states are affected by the gluonic 
quantum numbers \cite{hybrid}, which is in turn reminiscent of the fact that valence electrons modify the interaction between ions
in a non-trivial way in molecular physics. It would be important to have lattice estimates of $QQ$ potentials (static energies) 
for different quantum numbers of the light degrees of freedom. Finally, let me note that these
interesting systems are expected to be produced experimentally soon. In fact, a few states have already been reported \cite{exdhb}.  

\section{ Discussion}

The main issue at this conference is the understanding of the quark confinement 
mechanism, which turns it into a prominently theoretical event.
This is why I have put little emphasis on the experimental results. In order to partially compensate for it, let me just recall that most of the 
excitement in heavy quark physics is nowadays driven by experiment. Indeed experimental groups have had the larger
representation ever in the heavy quark session of this conference. This is not accidental. Many years after the discovery
of the first charmonium and bottomonium states, we see now close the time where the full spectrum below threshold will finally be 
uncovered, and precision data for many states is becoming available. Not to mention the wealth of results in $B$-physics among others. 

Having said that, let me come back to the links between heavy quark physics and
the confinement mechanism. In fact the goal of QCD based theory in heavy quark physics has been the 
parameterization of non-perturbative effects in a way as generic as possible so 
that any detail depending on the confinement mechanism is encoded in a number
of parameters (for instance NRQCD matrix elements) or functions (for instance the Isgur-Wise function). Since these parameters and functions can be extracted from experimental data,  heavy quark physics may help to discern between different
confinement mechanism in nature (e.g. monopoles \cite{suzuki} versus vortex \cite{faber}) if these mechanism provide different numbers or different shapes for the above mentioned parameters or functions.

In any case, progress in QCD based theory for heavy quarks goes on basically (but not only) along two lines: EFTs and improved lattice simulations. Still the use of 
phenomenological models is unavoidable to obtain estimates for many observables. I would like to emphasize that the cooperation 
between these three approaches may provide important results in the near future. 
For this cooperation to be 
successful, it is required that potential model practitioners try to up-grade their models to make them
 compatible with the current EFTs results and lattice QCD practitioners provide the non-perturbative QCD inputs that are
necessary to EFTs. 

Let me put an example to illustrate that this is an efficient way to proceed. Suppose we want to calculate the charmonium and bottomonium spectrum from QCD. Let us restrict for simplicity to vector and pseudoscalar states. If we insist in doing the calculation directly from the QCD lagrangian, we would have to calculate non-perturbatively four different two point functions, two of them with five dynamical quarks and the remaining two with four. If we instead use NRQCD at leading order, the calculation reduces considerably. Since spin symmetry tells us that pseudoscalars and vectors are roughly degenerated, we only have to calculate two two point functions with three dynamical quarks \cite{latticeNRQCD}. If we go further and use pNRQCD we only have to calculate the static potential with three dynamical quarks \footnote{I am assuming here that the $1/m$ potential is subleading.}, which is a much simpler non-perturbative calculation than any two point function either in QCD or NRQCD. The advantage of using pNRQCD becomes even clearer
if we wish next to do the same calculations for the $B_c$ system. For a QCD calculation we would have to calculate two extra two point functions, for NRQCD one extra two point function, and for pNRQCD nothing: the same static potential used for charmonium and bottomonium can be used for the $B_c$ system as well. Furthermore, since many potential models use potentials which are similar to the lattice
QCD static potential (e.g. the Cornell potential \cite{godfrey}) and reproduce data quite well, we are now close to be actually testing QCD.
 
Elaborating further on the proposal above, let me finally mention a few lattice calculations which are needed to make the link between potential models and QCD quantitative. First of all, the evaluation of the $1/m$ potential discovered in \cite{m1}. This potential has been ignored so far in potential models, but 
on general grounds it could be as important as the static potential. The phenomenological success of potential models (based on the static potential only) suggests that this will not be the case, but it should be confirmed by lattice data. Second, an updated evaluation of the spin dependent potentials \cite{gunar} would also be helpful. Finally, it has been shown in~\cite{pwave} that the inclusive decay widths require a few universal non-perturbative parameters, 
which should also be calculated on the lattice.

\section{Conclusions}

Let me just conclude with a few sentences which are  intended to flash what is going on in the field: (i) there is lot of excitement in experiment, (ii) progress from QCD persists, mainly through the use of EFT techniques and improved lattice simulations, and (iii) a great deal of modeling is still needed for many observables.

\section*{Acknowledgments}

I am grateful to Geoffrey Bodwin and Dieter Gromes for suggestions on the preparation of the talk, and to Antonio Pineda for a careful reading of this written version of it. \uppercase{T}his work is partially supported by \uppercase{M}y\uppercase{CT} and \uppercase{F}eder, \uppercase{FPA}2001-3598,
and by  \uppercase{CIRIT}, 2001\uppercase{SGR}-00065.

\end{document}